\documentclass[aps,twocolumn,superscriptaddress,showpacs,eqsecnum,nofootinbib,pre]{revtex4} 

\usepackage{scrextend}
\usepackage{amssymb}  
\usepackage{amsmath}   
\usepackage{natbib}
\usepackage{epsfig}   
\usepackage{relsize}
\usepackage{graphicx,subfigure}   
\usepackage{dcolumn}
\usepackage{bm}  
\usepackage{soul}
\usepackage[english]{babel}
\usepackage[latin1]{inputenc}
\usepackage{hyperref}
\usepackage{xcolor}
\usepackage{comment}
\usepackage[update]{epstopdf}
\begin{document}

\title{Population dynamics in an intermittent refuge}

 \author{E. H.  Colombo} \email{colombo@aluno.puc-rio.br}
\affiliation{Departament of Physics, PUC-Rio, Rio de Janeiro, Brazil}
\author{C. Anteneodo} \email{celia.fis@puc-rio.br}
\affiliation{Departament of Physics, PUC-Rio, Rio de Janeiro, Brazil}
\affiliation{Institute of Science and Technology for Complex Systems, Rio de Janeiro, Brazil}

\begin{abstract}
Population dynamics is constrained by the environment, which needs to obey 
certain conditions to support population growth. 
We consider a standard model for the evolution of
a single species population density, that includes reproduction, competition for resources and spatial spreading,  
while subject to an external harmful effect. 
The habitat is spatially heterogeneous, there existing a refuge  where the population can be protected. 
Temporal variability is introduced by the intermittent character of the refuge.
This scenario can apply to a wide range of situations, 
from  a lab setting where bacteria can be protected by a blinking mask from ultraviolet radiation, 
to  large scale ecosystems, like a marine reserve where there can be  seasonal fishing prohibitions.
Using analytical and numerical tools, we investigate the asymptotic behavior of the total population  
as a function of the size and characteristic time scales of the refuge.
We obtain expressions for the minimal size required for population survival, in the slow and fast time scale limits. 

\end{abstract}


\maketitle

\section{Introduction}

The collective behavior of living beings has been addressed in the literature 
by means of theoretical models combined with experimental observations, 
from microscopic to  ecological scales~\cite{active,turchin,cavagnaBirds,fish}. 
Besides the accomplishments in understanding the interaction between biological entities,  
the intrinsic role of the environment  in population dynamics has been discussed, 
as far as it has a critical impact in population survival~\cite{HanskiBook}. 
In order to predict the future of a given population in a particular habitat, 
it is necessary to take into account the nontrivial spatial distribution of resources, 
shelter, nutrients and other factors that compose the so called ecological landscape~\cite{landscapeBook}. 
Moreover, the ecological factors change in time with a characteristic periodicity (seasonality) accompanied by random fluctuations. 
Then, the environment critical conditions for population survival rely on a combination of the spatial and temporal variability  
of the environment~\cite{natureHanski,simon,ephemeral,OvaskCorrelated,nos3}.

A region, like a shelter, shield, mask, etc., 
that allows individuals to be protected against unfavorable conditions 
(e.g. predation, water scarcity, sun light)~\cite{landscapeBook} constitutes a refuge. 
Its time variability can have different origins. For instance, 
when the system is found in a natural habitat, it is typically subjected to inherent cycles of the ecosystem such as oscillations in sun light, seasonal changes, and other external dynamics that can interfere in the refuge conditions. 
For ecological reserves, where there exists certain control of the system \cite{marineHarvesting}, 
the time scale can be introduced, for instance, by fishing prohibition laws that are made flexible during specific periods of the year. 
In the case of microorganisms, where artificially constructed landscapes can be made~\cite{perry,pnasNano}, 
a time scale might be introduced in the experimental setup via manipulation of a mask that can protect 
a population of bacteria from a a harmful effect.
 
\begin{figure}[h]
\includegraphics[width=0.8\columnwidth]{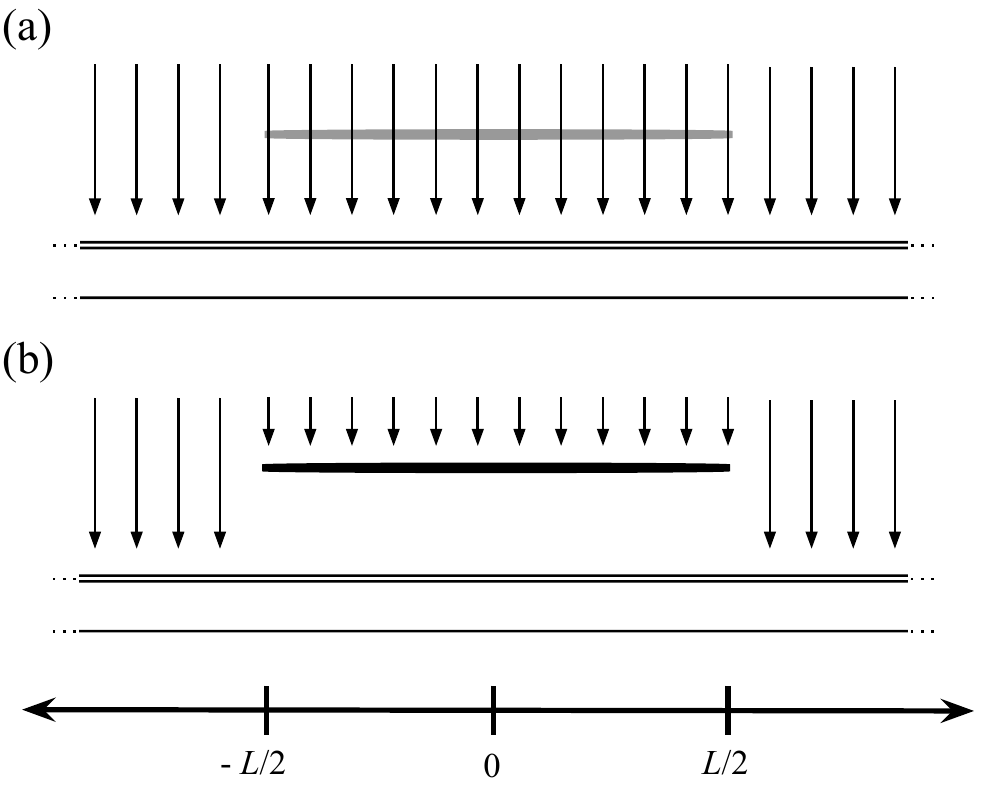}
\caption{Pictorial representation of a onedimensional habitat 
subject to an external harmful effect  (downwards arrows)  
with a refuge (thick segment) of size $L$ in the inactive (a) and active (b) states. 
In the active state, the refuge is able to block the harmful effect. 
}
\label{fig:setup}
\end{figure}

Changes in size~\cite{periodicKenkre}, position~\cite{walkmask} or even 
rotations~\cite{convectionPRE} of the refuge, as well as stochastic fluctuations~\cite{Horsthemke,oasis} 
have been considered before.
In this work, making basic assumptions about the population dynamics, 
we investigate population survival when there is an intermittent refuge of size $L$  (see Fig.~\ref{fig:setup}). 
We consider that the refuge alternates,  with period $\tau$, 
between  active and inactive states, such that the population can be protected or not, 
during intervals $\lambda\tau$ and $(1-\lambda)\tau$ (where $0\le \lambda \le 1$), respectively. 
We mainly investigate the requirements for survival  as a function of  the characteristic 
time scales  and size of the refuge,   
aiming to provide  general insights that can guide population management and conservation~\cite{partial}.

The mathematical model is defined in Sec.~\ref{sec:pop}.  
In Sec.~\ref{sec:static}, we review the static refuge case. 
In sec.~\ref{sec:slowfast}, we study the behavior of the population as a function of the $\tau$ and $\lambda$ 
when the refuge periodically alternates between the active and inactive states. 
Analytical expressions are obtained for the slow and fast limits. 
Details about the spatial dynamics are presented in Secs.~\ref{sec:spatial} and \ref{sec:recol}. 
Sec.~\ref{sec:final} contains final considerations.

\section{Population dynamics}
\label{sec:pop}

The temporal evolution of the population density distribution $u(x,t)$  is described by the 
Fisher-KPP equation~\cite{fisher1937wave,murray2003mathematical,murray2002mathematical} plus an additional term, 
namely,
\begin{equation}
\partial_t u(x,t) = D \partial_{xx}  u(x,t) + f(u) + \psi(x,t)u(x,t)\, , 
\label{maineq}
\end{equation}
where $D$ is the diffusion coefficient, 
 $f(u)$ is the local growth rate given by the logistic or Verhulst expression $f(u)=a u(x,t)\left[1 - \frac{u(x,t)}{K}  \right]$, 
with intrinsic growth rate $a$ and    carrying capacity   $K$, 
which bounds population growth, inducing negative growth rates for $u>K$.
For $\psi(x,t)=0$, one recovers the standard Fisher-KPP equation. 
In our model,  
\begin{equation}
\psi(x,t) =  -A \left[ 1- \Theta(L/2-|x|)\varphi(t) \right]\,,   
\label{eqpsi}
\end{equation}
with $A>a$. 
It contains the environment structure, pictorially represented in Fig.~\ref{fig:setup}, 
where a harmful effect is always present, contributing with an additional death rate in Eq.~(\ref{maineq}), 
but a refuge  located at  $|x|\leq L/2$ can mitigate the effect. 
The factor $\varphi(t)$ embodies the time variability of the refuge. 
If the refuge is absent or inactive (Fig.~\ref{fig:setup}.a), $\varphi(t)=0$, then  $\psi(x,t)=-A$ for all $x$. 
The  refuge can protect the region  $|x|\leq L/2$ (Fig.~\ref{fig:setup}.b),  
either partially (when $0<\varphi(t)<1$) or totally (when $\varphi(t)=1$).
For simplicity, we assume a binary time behavior, such that $\varphi(t)$ can only take the values  0 and 1. 
Additional parameters $\lambda$ and $\tau$ control the fraction of time that the harmful effect penetrates 
the refuge and the protocol time scale, respectively. 
Namely, during an interval $\lambda\tau$, the refuge is inactive, allowing the harmful effect to penetrate the refuge. Afterwards, the refuge becomes active, protecting the population during an interval $(1-\lambda)\tau$. 

Equation~(\ref{maineq}) will be numerically integrated by means of a standard 
fourth-order Runge-Kutta algorithm, together with spatial discretization, 
using $\Delta x=10^{-2}$  and $\Delta t<10^{-5}$, adequate for convergence.  
Along this work,  we will  focus mainly on population preservation at long times 
as a function of the refuge size $L$ and time scale $\tau$, keeping the remaining parameters fixed. 
Motivated by experiments for a nonchemotactic strain of \emph{E. Coli} bacteria~\cite{perry}, 
we set $a=1$, $K=10^{4}$, $D=10^{-1}$, $A = 6$, except when different values are explicitly indicated. 
Nevertheless, analytical expressions allow to extend the numerical results shown for that set of values.

\subsection{Static refuge case}
\label{sec:static}

\begin{figure}[h]
\includegraphics[width=\columnwidth]{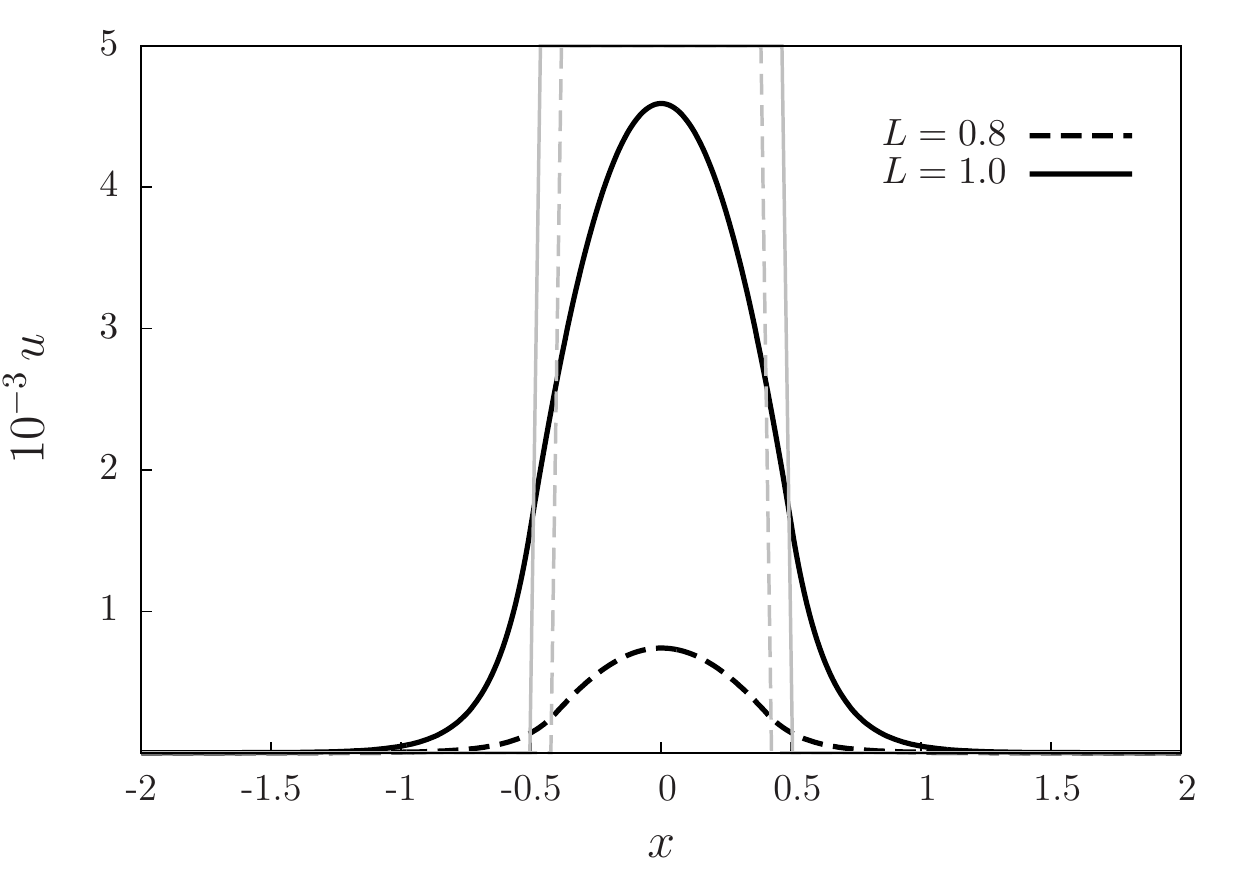}
\caption{Asymptotic population distribution when the refuge is static, 
for two different refuge sizes $L$ indicated in the figure. 
In this case, $L_c\simeq 0.73$, for our choice of the parameter values defined in Sec.~\ref{sec:pop}. 
The gray vertical lines indicate the boundaries of the refuge for each refuge size.
}
\label{fig:distributions}
\end{figure}

The case where the refuge is always active ($\varphi=1$) is well known in the literature \cite{skellam,ludwing}.
The refuge imposes an heterogeneous spatial condition which is the spatial component of $\psi$. 
When the refuge 
has size $L$ larger than a critical value $L_c$, the population survives achieving a nontrivial steady state. 
In Fig.~\ref{fig:distributions} we show the distribution profiles for two values of $L$, with $L>L_c$.

The critical refuge size $L_c$ can be obtained as follows. 
Assuming that there is a critical size $L_c$ at which the null solution of 
Eq.~(\ref{maineq}) becomes unstable, at that point, we can assume $u\simeq 0$ and 
consider the linear form of the equation. In this approximation, 
the nontrivial solution is trigonometric (exponential) inside (outside) the refuge. 
Imposing the continuity of that solution and its first order derivative at the refuge boundary, 
it is possible to find the value of the critical size $L_c$.  
Following this procedure, it is straightforward to obtain~\cite{ludwing,skellam,Kenkre2003,Kenkre2008}

\begin{equation}
L_c = L^* \equiv 2 \sqrt{\frac{D}{a}}\arctan \left(\sqrt{\frac{A-a}{a}}\right) \,.
\label{deterministic}
\end{equation}
In the literature, this result has been  extended to modified forms of the static Eq.~(\ref{maineq}), 
including advection, nonlinear diffusion, other boundary conditions and 
functional forms of $f$~\cite{skellam,ludwing,partial,Kenkre2003,advection,kraenkel}. 
For instance, in the limit of harsh unfavorable conditions,  $A \gg a $, Eq.~(\ref{deterministic}) 
yields $L_c \propto \sqrt{D/a}$ \cite{Kenkre2003}. 
For other cases, Eq.~(\ref{deterministic}) still holds for effective values of 
the rates inside and outside the refuge~\cite{partial,advection}.  
It is still a good reference even when demographic noise is included  to account for 
the fact that the population is constituted by a finite number of individuals~\cite{oasis}.

\section{RESULTS}
\label{sec:results}

In this section, we present our results that show the influence of refuge temporal variability in population conservation.  
We consider a refuge  whose temporal behavior is deterministic and periodic with period $\tau$.

\begin{figure}[h]
\includegraphics[width=\columnwidth]{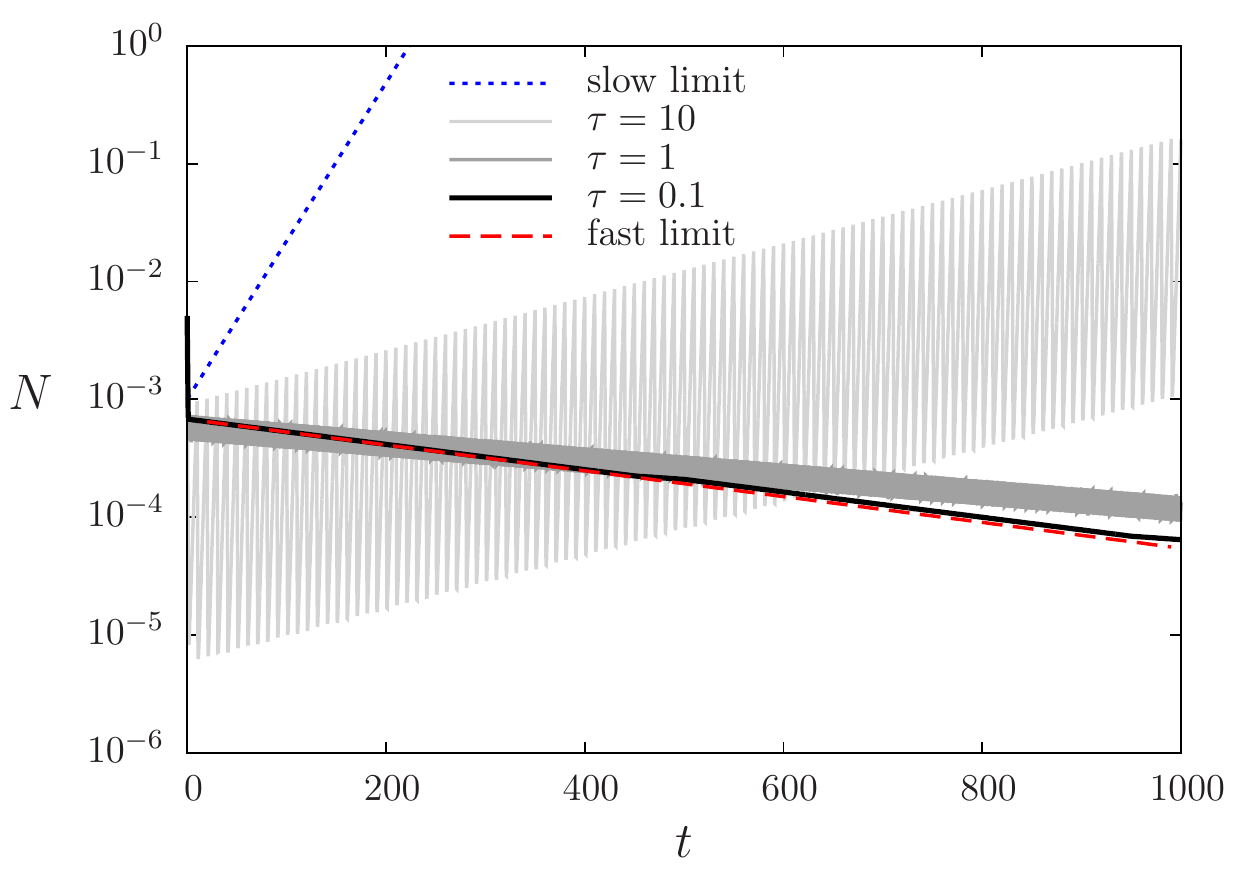}
\caption{Temporal evolution of the total population size $N$, for different values of the  period $\tau$,  
fixed average rate $\lambda=0.1$ and size $L=1.28$  
(for the values of the parameters used, $L_c = 1.295$).
The dotted and dashed lines represent  the slow  and fast limit approximation 
(for details, see Sec.~\ref{sec:slowfast}).}
\label{fig:near}
\end{figure}

\begin{figure}[b]
\includegraphics[width=\columnwidth]{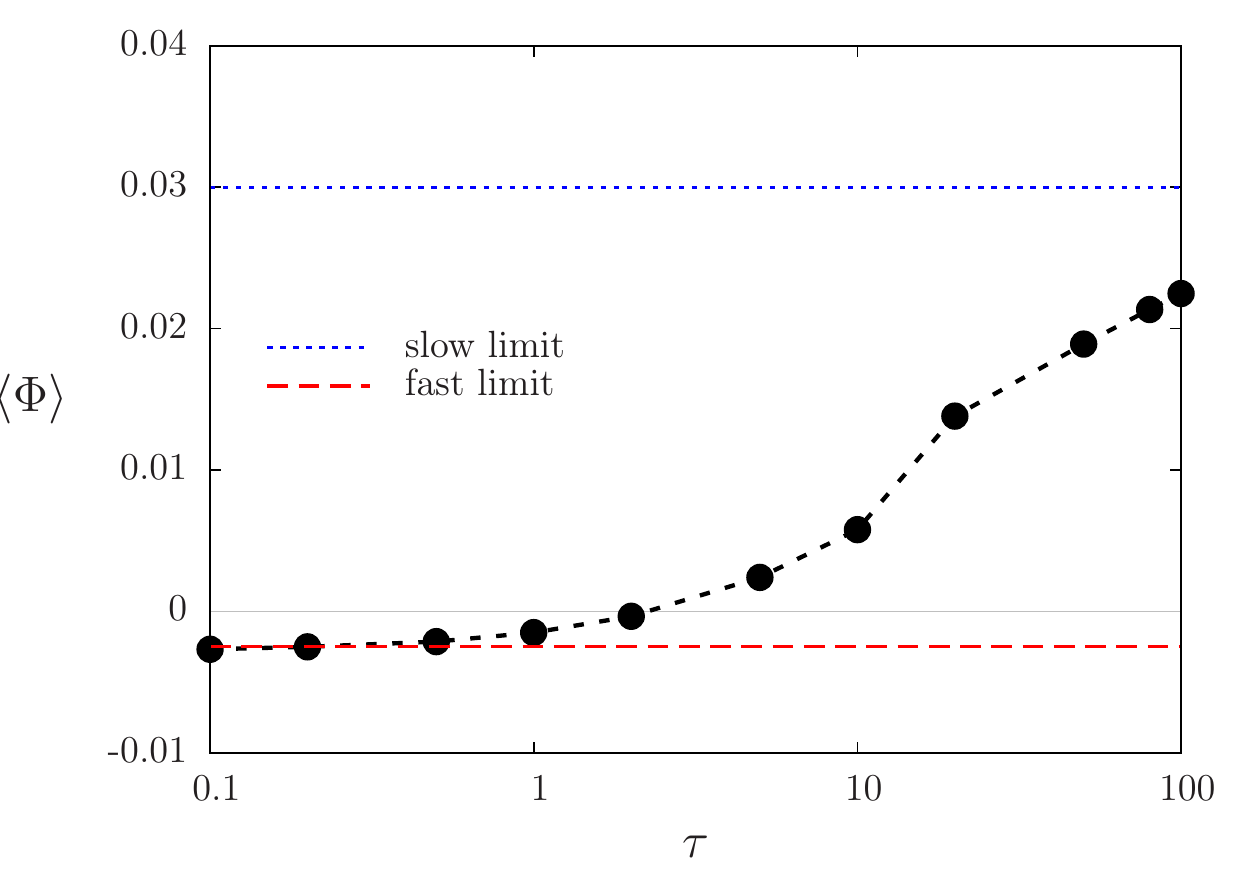}
\caption{Average population growth rate $\langle \Phi \rangle$, computed over entire cycles, 
as a function of protocol period $\tau$ for $L=1.28$ and $\lambda=0.1$. 
The dashed and dotted lines represent the rates at the slow  and fast  limits (for details, see Sec.~\ref{sec:slowfast}).  
}
\label{fig:growthtau}
\end{figure}

Figure~(\ref{fig:near})  shows the temporal evolution of the total population 
size $N(t)=\int_{-\infty}^{\infty} u(x,t) dx$, starting from population densities well 
below the carrying capacity ($u(x,0) \ll K$, for all $x$).  
We vary the time scale $\tau$ for fixed $\lambda$. 
Even if the fraction of time that the harmful effect penetrates the refuge is the same, 
we observe that, when subject to a fast varying environment, the population decays to extinction, 
but, differently, for large $\tau$, the population grows and survives at long times. 
This drastic change from extinction to survival  occurs because $L$ is near enough a critical, as we will see 
in  subsection \ref{sec:slowfast}. 
However, increasing $\tau$ favors population growth for any $L$.  
In order to show these effects, we define the growth rate per capita  
\begin{equation}
\Phi \equiv  \dot{N}/N = \frac{d (\ln N)}{dt} \,,
\end{equation}
whose average over one cycle is $\langle \Phi \rangle(t) = \frac{1}{\tau}\int_t^{t+\tau} \Phi(t')dt'$. 
After a short transient, while the population still remains low, this average attains a quasi-steady value 
$\langle \Phi \rangle$, 
corresponding to the average slope of the curves plotted in Fig.~(\ref{fig:near}). 
For negative $\langle \Phi \rangle$, its steady value will remain for long times, 
otherwise,  it will decay at later times when the population becomes comparable to the carrying 
capacity and stops growing attaining  a steady level.  
In Fig.~\ref{fig:growthtau}, we show  $\langle \Phi \rangle$  as a function of $\tau$. 
 
For the extreme cases of  slow and fast time scales, we show, in Sec.~\ref{sec:slowfast}, 
the derivation of the average growth rates, represented in 
Figs.~\ref{fig:near} and ~\ref{fig:growthtau}.  
These limits provide the bounds of the influence of refuge temporal variability.

\subsection{Slow and fast limits}
\label{sec:slowfast}

First, we start by assuming that  flashes occur in a very short time scale $\tau \ll \tau_S = 1/a$, 
such that the system does not have time to respond, where $\tau_S$ is the system time scale. 
In this limit, environment fluctuations can be locally averaged, 
producing an effective growth inside the refuge $(1-\lambda)a - \lambda (A-a)= a-\lambda A$ 
(dashed line in Fig.~\ref{fig:near}). 
Substituting the intrinsic growth rate $a$ by the effective one into Eq.~(\ref{deterministic}), gives 

\begin{equation}
L_c(\lambda;\tau\ll\tau_S) = 2 \sqrt{\frac{D}{a-A\lambda}}\arctan \left(\sqrt{\frac{A-a}{a-A\lambda}}\right) \, ,
\label{critfast}
\end{equation}
where $\tau_S \sim 1/a$ is the system response time. 
This result is expected to be independent on the microscopic details of the protocol, i.e. 
whether it is regular or stochastic behavior, 
being only dependent on its averaged behavior, characterized by parameter $\lambda$.

\begin{figure}[t]
\includegraphics[width=\columnwidth]{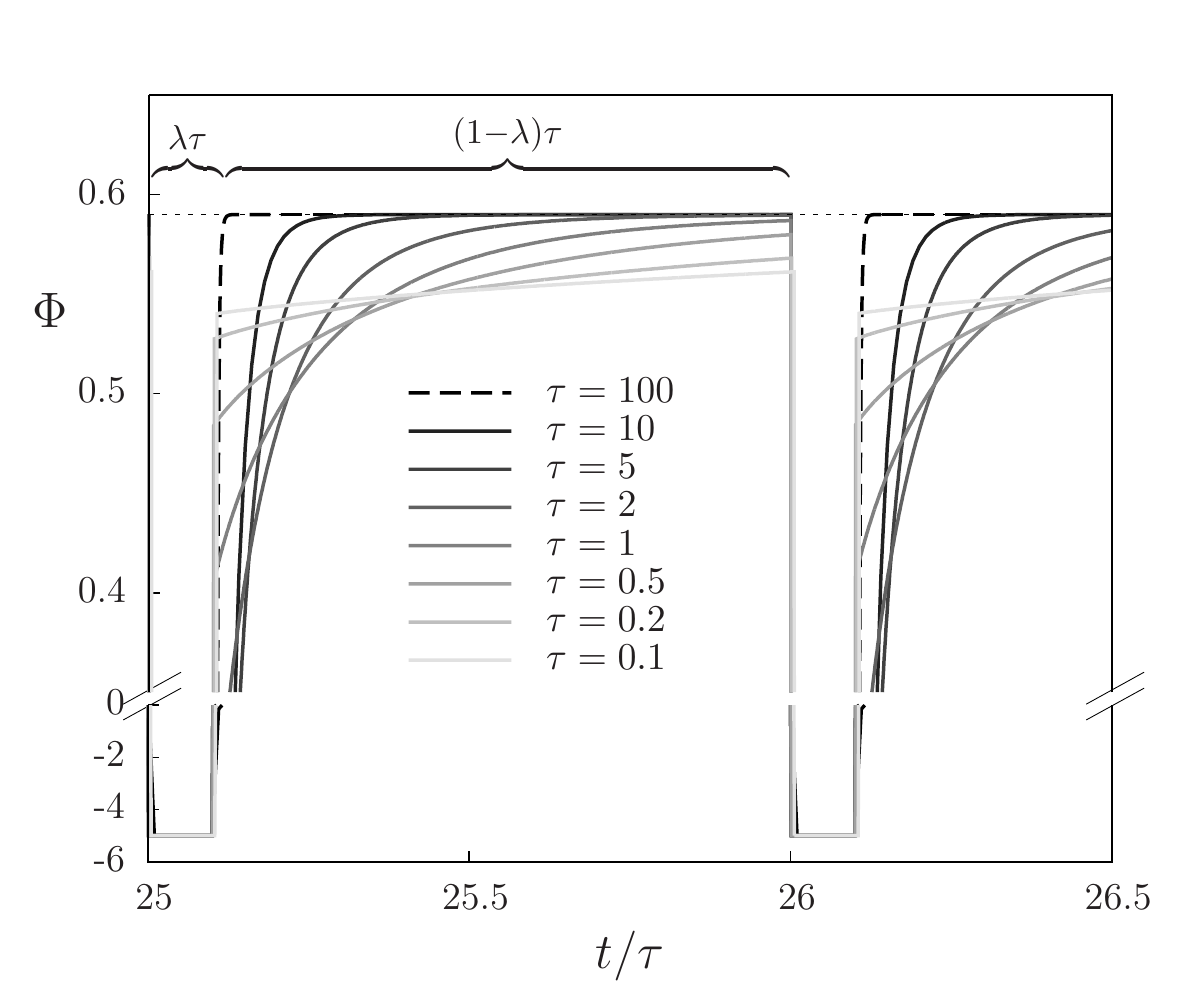}
\caption{Temporal evolution of the growth rate $\Phi$, for different time scales $\tau$, with $L=1.28$ and $\lambda=0.1$. 
The dotted line denotes the maximal value $\Phi_0$. }
\label{fig:relativegrowth}
\end{figure}

In order to estimate the slow-limit behavior, it is useful to observe the evolution of the growth rate $\Phi$, 
for different time scales $\tau$, 
as depicted in Fig.~\ref{fig:relativegrowth}, where we have rescaled time $t$ to facilitate the comparison of different periods $\tau$. 
During the interval $\lambda \tau$, when the harmful effect penetrates the refuge, 
the growth is negative, constant and independent of time scale. 
When the harmful effect is blocked, the growth rate tends to attain a maximal value $\Phi_0$, which is achieved for large $\tau$, $\tau \gg \tau_S$.

\begin{figure}[b]
\includegraphics[width=\columnwidth]{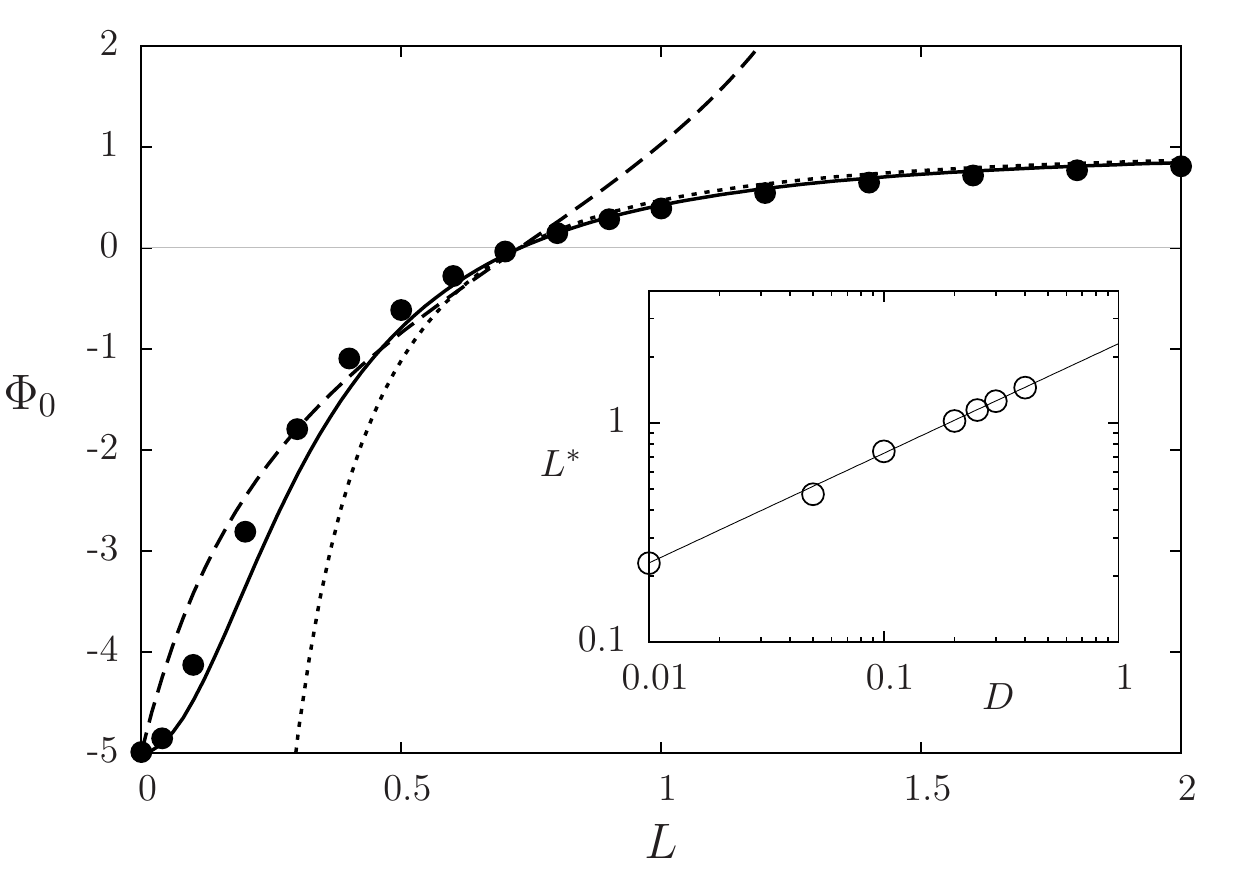}
\caption{Population growth rate $\Phi_0$ vs  refuge size $L$ (black circles). 
The solid line represents the ansatz given by Eq.~(\ref{ansatz}), and the 
dotted line represents the growth rate in the limit case of harsh conditions outside the refuge, explicitly given by Eq.~\ref{harsh}, 
and the dashed line the linear approximation given by Eq.~\ref{Phi0}. 
In the inset, we show that the fitting parameter $L^*$ in the ansatz (\ref{ansatz}) follows  Eq.~(\ref{deterministic}).
}
\label{fig:restore}
\end{figure}

In this slow limit, we  approximate the average growth rate by 
$\langle \Phi \rangle \approx (1-\lambda)\Phi_0(L) - \lambda (A-a)$. 
Then, imposing $\langle \Phi \rangle = 0$, the critical refuge size under slow environmental changes can be written 
by using the inverse function of the growth rate, 
$L_c \simeq \Phi_0^{(-1)}[\lambda(A-a)/(1-\lambda)]$.

The behavior of  $\Phi_0$ as a function of refuge size is shown in Fig.~\ref{fig:restore}. 
Approximate expressions for $\Phi_0(L)$ are presented in appendix~\ref{sec:appendixPhi}. 
The numerical data can be well described by the heuristic expression  (see appendix \ref{sec:appendixPhi}) 
\begin{equation} \label{ansatz}
\Phi_0(L) = a - \frac{A}{1 + \frac{A-a}{a}(L/L^*)^2}\, , 
\end{equation}
where $L^\star$ is the static case critical size given by Eq.~(\ref{deterministic}). 
Explicitly, the critical refuge size for the slow limit becomes

\begin{equation}
\label{critslow}
 L_c(\lambda;\tau\gg \tau_S) = L^*\sqrt { {\frac {a}{   a -\lambda A   }}}\, .
\end{equation}

\begin{figure}[h]
\includegraphics[width=\columnwidth]{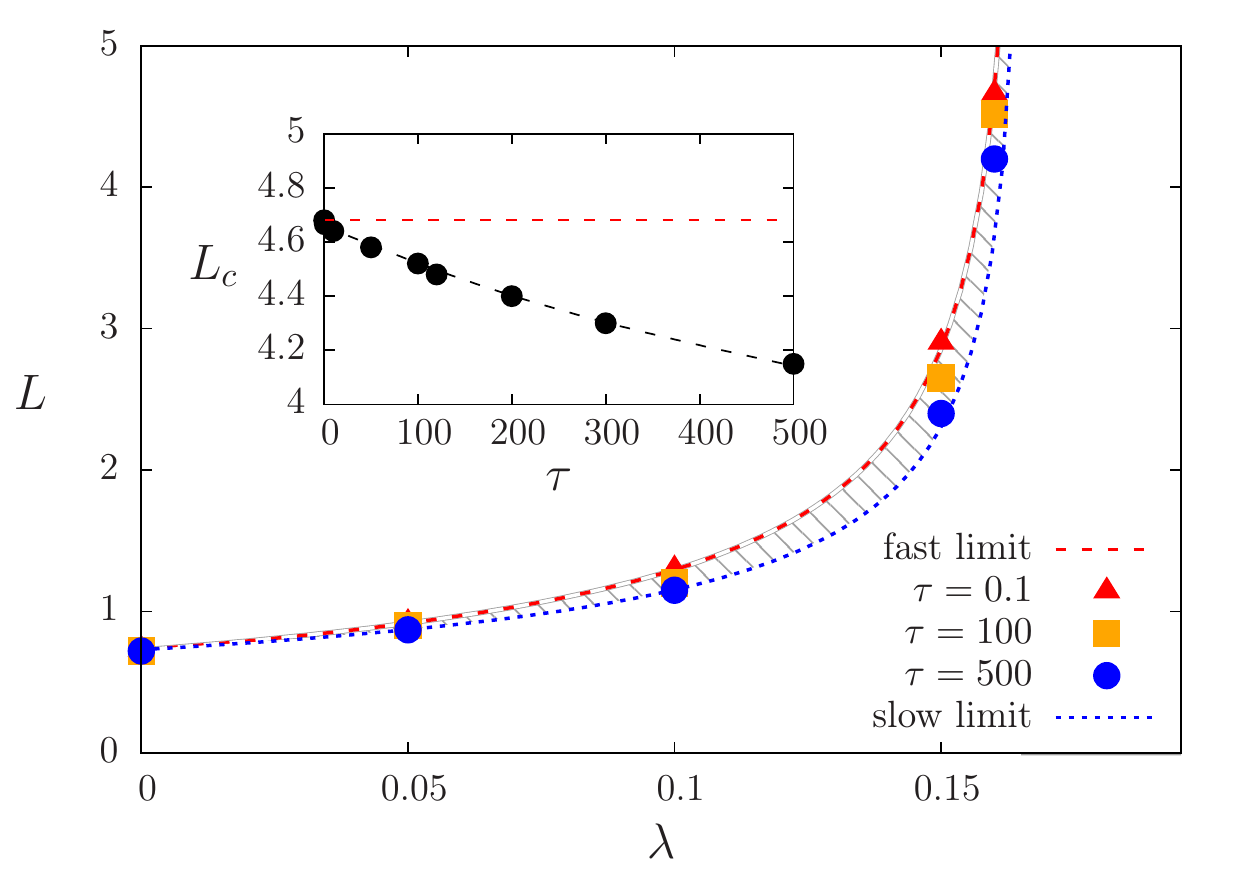}
\caption{Theoretical predictions for the critical size $L_c$ in the slow and fast protocol limits, 
given by Eqs.~(\ref{critslow}) and (\ref{critfast}), respectively, 
together with numerical data for different $\tau$. 
The stripped region (black) between the curves represents the variability of  $L_c$ with $\tau$. 
The inset shows $L_c$ vs $\tau$ for $\lambda=0.16$ and the slow limit approximation (blue dotted line).  
}
\label{fig:critical}
\end{figure}

We summarize the results of this section in Fig.~\ref{fig:critical}, 
where we show the upper and lower bounds for the critical size 
$L_c(\lambda;\tau\gg \tau_S) \leq L_c \leq L_c(\lambda;\tau\ll \tau_S)$, 
together with numerical results for different values of $\tau$. 
The dashed region represents the possible range of $L_c$ as a function of protocol temporal behavior.
A critical value of $\tau$, for which the average 
growth rate $\langle \Phi \rangle$ changes sign, always exists for $L$ within that range.

Notice that when $\lambda=0$, the bounds given by Eqs.~\ref{critfast} and \ref{critslow} coincide, 
recovering the static value of $L_c$. 
In the limit $\lambda \to a/A<1$, the critical size is divergent.

\subsection{Spatial dynamics}
\label{sec:spatial}

In this section, we will focus on the mechanism that connects the spatial and temporal components of the environment.
Considering the low density regime $u \ll K$, and integrating Eq.~(\ref{maineq}) in space, we obtain that

\begin{align}
\partial_t N = -(A-a)N_{out} + \{[a-A[1-\varphi(t)]\}N_{in}
\label{inout}
\end{align}
where $N_{\text{in}}$ and $N_{\text{out}}$ are the total populations inside and outside the refuge domain, 
respectively. 
Due to the fact that population growth occurs only inside the refuge, the external population  is  the result of 
the accumulated flux of individuals leaving the refuge. 
This makes the unfavorable neighborhood work as a reservoir of individuals. 
Explicitly, in the linear regime, 
$\dot N_{\text{out}}(t) = -(A-a)N_{\text{out}} + J$, 
where $J/D= -2(\pm\partial u/\partial x)|_{x=\pm L/2} = 2(Vu)|_{x= \pm L/2}$  is the flux through the refuge boundary and $ V$ the net velocity outward the refuge.   
The equation for the population inside the refuge is simply 
$\dot{N}_{\text{in}} = \{[a-A[1-\varphi(t)]\}N_{in} - J$. 
Due to the combination of a nonlinear spatial dynamics and heterogeneous environment, 
the flux $J$ has a nonlinear dependency with $N_{\text{in}}$ and $N_{\text{out}}$ and 
it is also history-dependent. 
This means that attempts to define $J$ as proportional to the population density difference 
$N_{\text{in}} - N_{\text{out}}$ ignore  nonlinearities  of the spatial dynamics and will not be 
suitable to model the system behavior  (see Sec.~\ref{sec:recol}), 
yielding  $\tau$-independent results.

\begin{figure}[b]
\includegraphics[width=0.9\columnwidth]{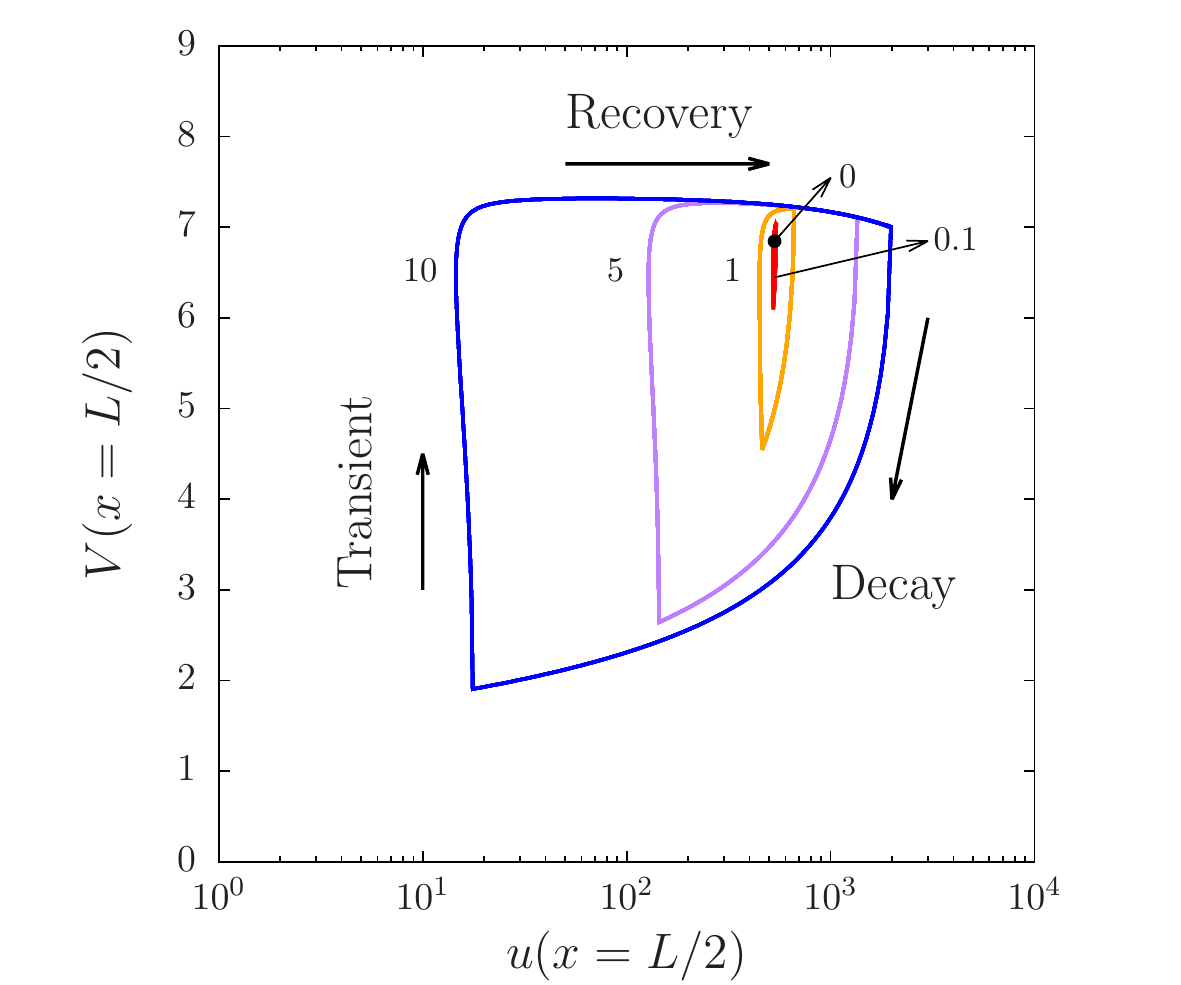}
\caption{Velocity $V(x=L/2)$ vs population density $u(x=L/2)$ at the refuge boundary,  
for different values of $\tau$ indicated in the figure, with  $\lambda=0.1$  and  refuge size $L=2$ (hence, $L>L_c$). 
The single dot represents the limiting case $\tau\to 0$.
}
\label{fig:flux}
\end{figure}

For the case of a time periodic protocol, in Fig.~\ref{fig:flux}, 
we show typical trajectories in the plane $u-V$, where the density and the velocity are evaluated  at one of the boundaries. 
Time integration of these trajectories provides the total flux that left the refuge.  
The emergent cycles are induced by the protocol and their shape reveals the relation between 
the localized perturbation produced by the protocol and spatial changes in population distribution. 
First, when the condition inside the refuge changes from favorable to unfavorable, 
the population decays and its population tends to be flattened, 
as we see in Fig.~\ref{fig:flux}, $V(x=L/2)$ decreases (decay period). 
When the refuge becomes active, the population inside the refuge starts to grow while the surrounding population 
is in constant process of extinction. 
This creates a fast stretch of the distribution, rapidly increasing the derivative 
of the population distribution at the refuge boundary (transient period). 
After the transient,   relaxation towards the steady state occurs, 
where the velocity at the boundary is kept roughly constant (recovery period), 
$|V| = \sqrt{(A-a)/D} \approx 7.0$ in the case of the figure, 
as predicted by the linear approximation (see Appendix \ref{sec:appendixPhi}).

As shown in Fig.~\ref{fig:flux}, for $L>L_c$, in the steady state, these cycles are closed curves. 
In contrast, for $L<L_c$, the curves are not closed, although the shape drawn in Fig.~\ref{fig:flux} remains essentially the same, 
but, at each period, the cycle is progressively shifted to the left (i.e., towards lower densities).

\subsection{Recolonization process}
\label{sec:recol}

\begin{figure}[h]
\includegraphics[width=\columnwidth]{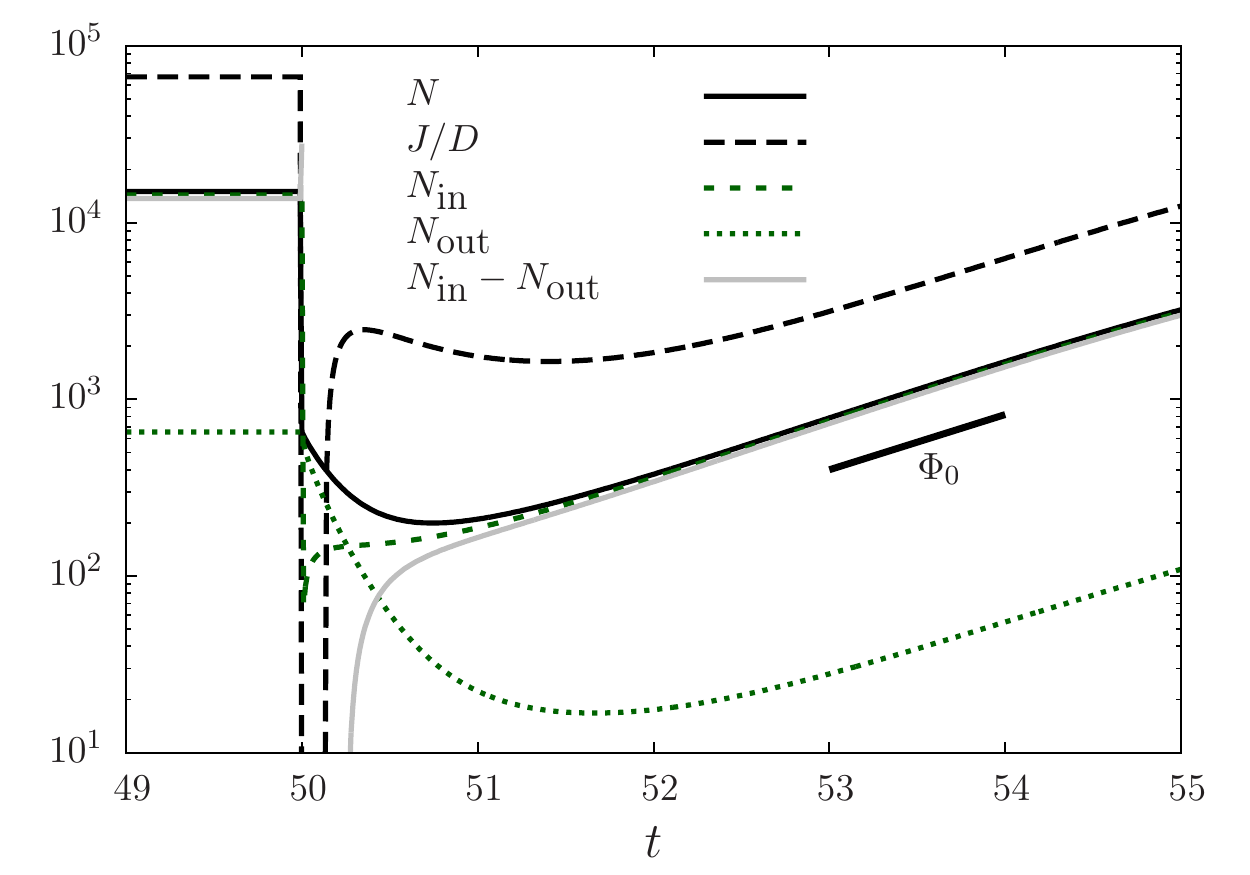}
\caption{Temporal evolution of the total population $N$, 
the scaled flux $J/D$, 
the populations inside  $N_{\text{in}}$ and  outside $N_{\text{out}}$ the refuge  
(after resetting the population inside the refuge at $t=50$) and their difference. 
In this case, the refuge size  is $L=2>L_c$. 
After the transient, the population achieves the recovery state, growing with rate $\Phi_0$.
}
\label{fig:reset}
\end{figure}

We now proceed to investigate the recolonization process that occurs when 
all the population inside the refuge dies due to a catastrophic phenomenon. 
Such extreme situation allows us to follow in detail 
the recolonization process that takes place from the lateral population reservoirs formed 
during the period when the refuge was active.
In Fig.~\ref{fig:reset}, we show, for $L>L_c$, how the total  flux at the borders and 
the population densities inside and outside the refuge behave during the recolonization process.
Focusing on the temporal evolution of $N_{\text{in}}$, it is very clear that the population growth 
is maximal just after the reset ($t=50$). 
This occurs due to the migration of the `stocked' population in the vicinity of the refuge.
This is confirmed by the change in the sign of the flux $J$, which becomes negative just after 
the reset, indicating that the net flux is inwards the refuge. 
Due to the fact that the source of the surrounding population is the flux of individuals from 
the refuge (see Eq.~(\ref{inout})), we can say that the environment spatial structure introduces 
a dependency on the history of the system. This is revealed by the non-monotonic response of 
$J$ and $N_{\text{in}}$ in Fig.~\ref{fig:reset}. Moreover, comparing the flux $J$ 
with the population difference $N_{\text{in}}-N_{\text{out}}$, it is clear that the simplification 
of  Eq.~(\ref{maineq}) to a two-population model~\cite{HanskiBook}, defining $J \propto 
(N_{\text{in}}-N_{\text{out}}) $ will not reproduce the observed behavior.

\section{Final considerations}
\label{sec:final}
 
We considered a refuge of size $L$ that periodically switches  between  active and inactive states, 
either protecting or not the population from a harmful external effect. 
We have investigated the critical spatiotemporal conditions for the conservation of a population 
in such intermittent refuge by means of numerical simulations.  

We provided analytical expressions for the critical refuge size $L_c$ 
at the slow and fast limits, which represent the lower and upper bounds for $L_c$, respectively. 
That is, in order to preserve the population, the refuge size in fast varying conditions needs 
to be larger when compared to the slow limit.
This means that for fixed  refuge size $L$ and the fraction of time $\lambda$ 
(that the harmful effect penetrates the refuge), 
the population growth is favored in a slowly varying environment (large $\tau$). 

In order to check the generality of these results, we also considered modified protocols. 
Instead of the binary case where $\varphi$ takes the values 0 and 1, 
we also used smooth periodic profiles varying continuously between 0 and 1, 
while keeping the integral fixed for comparison. 
Moreover, we also considered protocols with (uniform) random fluctuations in the duration 
of the active and inactive periods.
Implementing these protocols, we observed the same phenomenology that for the deterministic 
binary case described in detail in Sec. ~\ref{sec:results}, yielding results qualitatively similar 
to those shown in Figs.~\ref{fig:near} and ~\ref{fig:growthtau}. 
Furthermore, although generically there is  a quantitative dependency on the precise profile shape, 
 discrepancies become negligible in the fast limit (small $\tau$), 
depending only on the average $\lambda$, for the remaining parameters fixed. 

Our results may be interesting for  conservation and management in the context of ecological 
reserves~\cite{marineHarvesting,partial}, where temporal variability is a relevant factor.
Experimental tests might be performed for microorganisms~\cite{convectionPRE,perry}. 
For instance, in the static case $\varphi(t)=1$, in Ref.~\cite{perry}, 
the author provides an experimental setup to validate 
 Eq.~(\ref{maineq}) for the determination of the critical refuge size $L_c$ in bacterial populations. 
In that case, a refuge exists due to a mask that protects the bacteria  
from a harmful UV light field, similar to Fig.~\ref{fig:setup}.   
Following that setup, the validity of our results might be checked by the introducing 
the intermittent behavior of the refuge through the manipulation of the mask.

{\bf Acknowledgments:} C.A. and E.H.C. acknowledge the financial support of Brazilian
Research Agencies CNPq and FAPERJ.

\appendix

\section{Population growth in heterogeneous static environment}
\label{sec:appendixPhi}

Assuming that population density is low, such that we can neglect the second order term from the carrying capacity, 
 the temporal evolution of the population spatial distribution in Fourier space, $\tilde{u}$, from 
 Eq.~(\ref{maineq}),  is given by 
\begin{equation}
\partial_t \tilde{u}(k,t) = (-Dk^2 + a) \tilde{u}(k,t) + [\tilde{\psi}\star\tilde{u}](k)  \,,
\label{linfourier}
\end{equation}
where the symbol $\star$ denotes the convolution operation, i.e. 
$ \tilde{\psi}\star\tilde{u} = \int \tilde\psi(k-k',t)\tilde{u}(k',t) dk'$. 
From the protocol definition in Sec.~(\ref{sec:pop}),   we obtain  
$\tilde\psi(k,t) = -A\left[\delta(k) - 2\frac{\sin(kL/2)}{ k}\right]$, 
where we consider the static case, setting $\varphi(t)=1$ for all $t$. 
The growth rate of the total population size is obtained by taking $k=0$,

\begin{align}
\partial_t \tilde{u}(0,t) &= (a-A)\tilde{u}(0,t) + A\int_{-\infty}^{\infty} \frac{2\sin(kL/2)}{k}\tilde{u}(k,t)dk \,.
\label{growth0}
\end{align}

In Sec.~\ref{sec:spatial}, the analysis of the spatial dynamics has shown, among other results, 
that, when the population grows during the recovery time, the spatial distribution changes 
but preserving its shape  (see Fig.~\ref{fig:flux}). 
Therefore, we assume that $u(x,t)=N(t)u_s(x)$. 
Then,  $\tilde{u}(k,t)=\tilde{u}(0,t)\tilde{u}_s(k)$,  where we have  arbitrarily set $\tilde{u}_s(0)=1$.  
As a consequence, we can write Eq.~(\ref{growth0}) as $\partial_t \tilde{u}(0,t)  = \Phi_0(L)\tilde{u}(0,t) $, 
with the intrinsic population growth rate being
\begin{equation}
\Phi_0(L)= a + A[S(L)-1] \,,
\label{Phi0}
\end{equation}
where 
\begin{equation} \label{SL}
S(L)\equiv  \int_{-\infty}^{\infty} \frac{2\tilde{u}_s(k)\sin(kL/2)}{k}dk.
\end{equation} 

First, we see that, independently of the shape of the distribution $\tilde{u}_s$, 
if $L\to 0$, then $S(L)\to 0$, and as a consequence  $\Phi_0\to a-A$. 
Second, in the limit of large refuge $L\to\infty$, we have $\tilde{\psi} \to \delta(k)$, 
then $S(L) = 1$, giving $\Phi_0 = a$.

We proceed obtaining an approximate expression for the distribution $u_s(x)$.
We start by recalling the steady solution  of Eq.~(\ref{maineq}), in the static case, 
for $L=L_c=L^*$ (see Eq.~(\ref{deterministic})): 

\begin{equation}
u(x,t) =
\begin{cases} 
      c_1 \cos(\beta_+ x) & |x|\leq L_c/2 \\
      c_2 e^{-\beta_-|x|} & |x| > L_c/2 
   \end{cases}
\label{distribution}
\end{equation}
where the parameters that regulate the spatial scale are $\beta_+=\sqrt{a/D}$, 
$\beta_- = \sqrt{(A-a)/D}$ and the constants $c_1$ and $c_2$ are such that $u(x,t)$ 
is continuous and differentiable at $x=\pm L_c/2$.
 Eq.~(\ref{distribution}) can be used as a base to estimate the shape of the distribution in the recovery period, 
for other values of $L$. 
In order to do  that, we keep the simple form of the critical solution but  flexibilize  the conditions 
at the boundary of the refuge, allowing discontinuity of the first derivative. 
This yields $c_1 = c_2\exp(-L_c\beta_-/2)/\cos(\beta_+ L_c/2)$. 
Normalization of Eq.~(\ref{distribution}) provides  the value of $c_2(L)$ (expression not shown). 
Then, the Fourier transform $\tilde{u}_k$ can be computed and substituted into Eq.~(\ref{SL}),  
giving 
\begin{equation}
S(L) = c_2(L)\exp(\beta_-L/2)\tan(\beta_+L/2)/\beta_+ \,.
\label{ap1}
\end{equation}

This expression is exact for $L_c$, where $\Phi_0(L_c)=0$ and captures the main contributions 
for $L<L_c$,  since the presence of higher modes in the limit of small $L$ is filtered by the shape $\tilde\psi$.
For $L>L_c$, the trigonometric solution loses its validity and the distribution tends to flatten. For this case, 
small values of $k$ (long wavelenghts) have a significant impact on $S(L)$.
%
In order to provide an analytical expression for small and large values of $L$, we propose the suitable ansatz 
\begin{equation}
S(L)= 1 - \frac{1}{1+\frac{A-a}{a}(L/L^\star)^2}  \,,
\end{equation}
 therefore 
\begin{equation}
\Phi_0(L) = a - \frac{A}{1 + \frac{A-a}{a}(L/L^*)^2}\, , 
\label{choice}
\end{equation} 
where $L^*$ is the critical refuge size in the static case.  
The expression in Eq.~(\ref{choice}) recovers the known result for hash conditions 
when $A\to\infty$, the asymptotic behavior for large $L$, and the condition $\Phi_0(L_c) = 0$. 
Comparison between Eq.~(\ref{Phi0}) (assuming $S(L)$ as in Eq.~(\ref{ap1}) ), our proposal Eq.~(\ref{choice}) 
and numerical data is shown in Fig.~\ref{fig:restore}.

\section{Slow and fast limits with harsh conditions outside the refuge}
\label{sec:harsh}

In the limit of harsh conditions outside the refuge, 
the population density goes to zero at the refuge boundary, i.e. $u(|x|>L/2)=0$. 
Under this boundary condition, it is straightforward to obtain the largest eigenvalue which determines 
the value of the growth rate $\Phi_0$ \cite{skellam,advection}, 
\begin{equation} \label{harsh}
\Phi_0 = a - \frac{\pi^2 D}{L^2} \,.
\end{equation}
The condition $\Phi_0 =0$ gives  $L^* = \pi\sqrt{D/a}$.

Following the same procedure described in Sec.~(\ref{critslow}),
 in the fast limit, we assume that the growth rate is  locally averaged, then
\begin{equation}
L_c(\lambda;\tau \ll \tau_S)= \pi \sqrt{\frac{D}{a-\lambda A}} = L^* \sqrt{\frac{a}{a-\lambda A}}   \,.
\end{equation}

In the slow limit, we assume that population growth switches 
between $a-A<0$, during the harmful action, and $\Phi_0$, 
during the recovery period. 
Then, $\langle \Phi \rangle = (a-A)\lambda + \left(a-\frac{\pi^2D}{L^2}\right)(1-\lambda)$. 
When $\langle \Phi \rangle=0$,  hence $L=L_c$,  we find that 
\begin{equation}  
L_c(\lambda;\tau\gg \tau_S)  =  \pi \sqrt{ \frac{D(1-\lambda)}{a-\lambda A} } =
 L^* \sqrt{ \frac{a(1-\lambda)}{a-\lambda A} } \,.
\end{equation}

Therefore, the ratio between the critical refuge sizes in the slow and fast limits is
\begin{equation}  
\frac{L_c(\lambda;\tau \ll \tau_S)}{L_c(\lambda;\tau\gg \tau_S) } = \frac{1}{\sqrt{1-\lambda}}  
\,.
\end{equation}

This means that, even in this case, where we neglect the role of the surrounding population, 
the spatial dynamics distinguishes slow from fast environment perturbations. 
Nevertheless, the ratio is  only 
$\frac{L_c(\lambda;\tau \ll \tau_S)}{L_c(\lambda;\tau\gg \tau_S) } = 1.091$, for  
$\lambda = a/A$. 
Therefore,  there is a relative difference of about  $9\%$   in refuge 
critical size due to temporal variability of the environment. 
However, when  conditions are not harsh outside, like in the case of Fig.~\ref{fig:critical}, the change in 
$L_c$ with $\tau$ can reach $30\%$.


%

\end{document}